# On Timeless Macroscopic Spaces

*Günter Nimtz and Horst Aichmann**

*II. Physikalisches Institut, Universität zu Köln,*

**Zur Bitz 1, Bad Nauheim/Germany*

*E-mail: g.nimtz@uni-koeln.de, *h-aichmann@t-online.de*

We begin the Article with confusing citations in published papers on the question recently: “how much time does a wave packet spend in a tunnelling barrier?”

> ***..a particle tunnelling through a barrier appears to do so in zero time [1]. .. The pulse transit through the barrier itself seems to be instantaneous [2]. ..tunnelling is unlike to be an instantaneous process [3]. ..ionization time is close to zero [4]. ..all waves have a zero tunneling time [5]. ..Our results are inconsistent with claims that tunnelling takes zero time [6].***

Sommerfeld and other physicists pointed out that tunneling is no classical process [5]. The arguments are that the tunneling wave’s momentum is imaginary and its energy is negative. However, what is the time $\tau$ inside a barrier? Brillouin conjectured that the non-classical behavior is valid for waves of all fields, i.e. for electromagnetic and acoustic waves as well as for Schrödinger waves. The first zero time inside a barrier was observed with microwaves [2]. Later on acoustic experiments pointed to a zero time. Recently, it became popular to try to measure and calculate $\tau$ of Schrödinger waves in the attosecond time regime by the help of laser pulses [6, 7]. All authors didn’t cite nor scale up the nanosecond time tunneling data carried out with microwaves. This note is devoted to look at the 4th dimension, i.e. the time $\tau$ waves spent inside the tunnel – barrier. Microwave experiments showed that virtual waves can extent up to several meters in zero time.

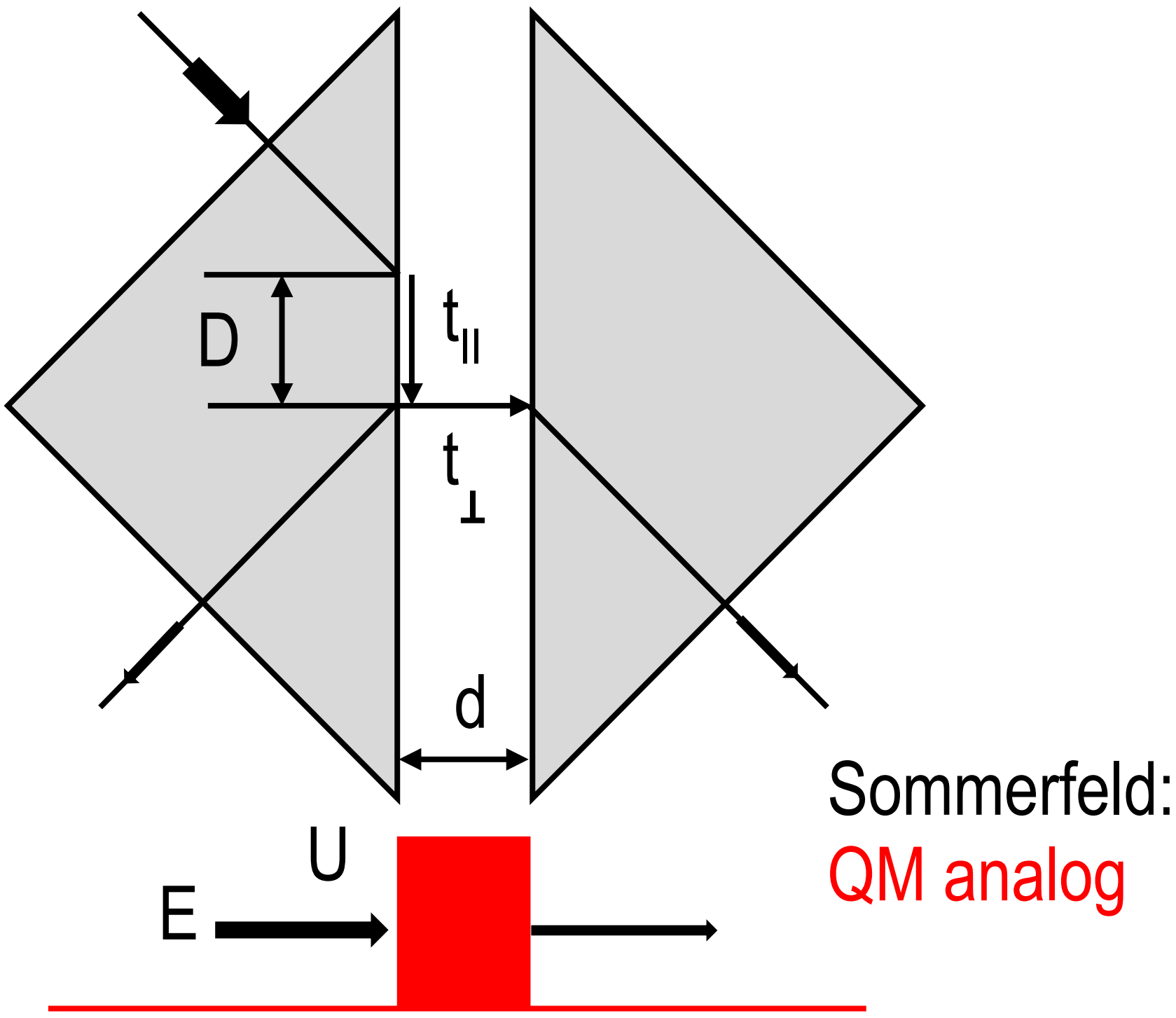

*Fig.1 Double prisms are displayed. The red potential barrier represents the barrier of the corresponding optical gap as studied in optics to measure the various time components. Incidentally technical components using the tunneling effect are applied in modern fiber optic communication systems in combiners and splitters.*

The tunneling set-up in Fig.1 is used to measure the so called frustrated total internal reflection. It is called frustrated since a small part of the total reflected wave is tunneling the gap to the second prism. It is a simple, well-defined setup, since the reflected and the transmitted waves differ in both, distance d and crossing time $\tau_{\perp}$ in the barrier. Thus, the crossing time can be measured.

A sophisticated quantum mechanical experiment was invented by C. Hong, Z. Ou, and L. Mandel [9] to measure the total photon velocity of single photons. This set-up was applied by A. Steinberg, P. Kwiat and R. Chiao [10], and they observed a superluminal single photon tunneling velocity. However, the measured superluminal velocity of the tunneled photon combines three individual time elements:

1. Entering the barrier, (incidentally, the reflection time equals the transmission time of barriers [11])
2. A potential time $\tau$ in the barrier, and
3. The time lost at the exit.

These authors spent much time with introducing strange models, for instance using the deformed shape of tortoises as a model for a wave packet in order to safe causality as displayed in Fig. 2.
.

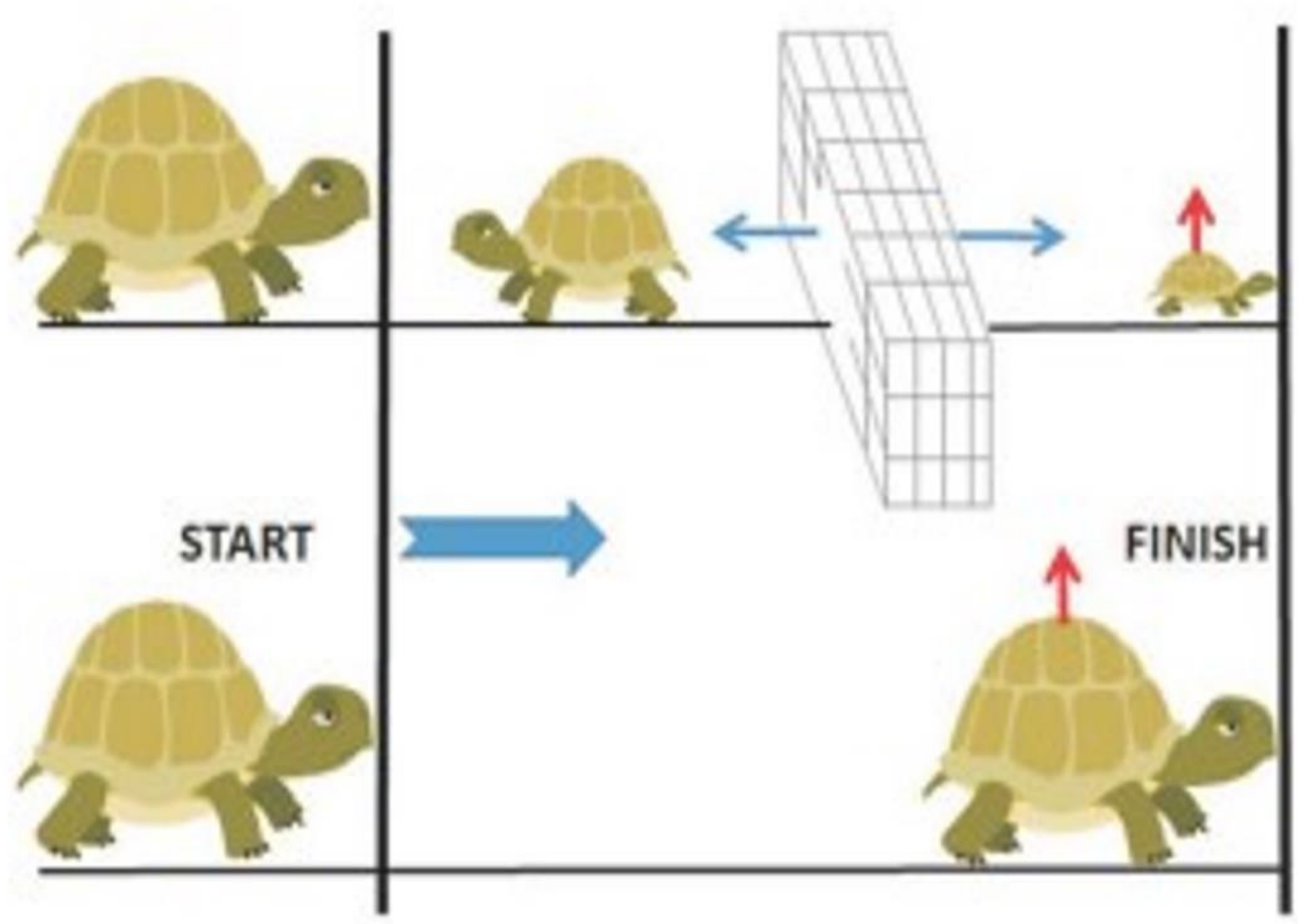

*Fig. 2 The front velocity after tunneling is luminal, whereas the center of mass (red arrows) has tunneled faster than light [12].*

Another example for not faster than light given by Steinberg is a train dropping the last wagon at each station stop in order to get the train's center of mass faster [13] There are many studies published in order to save causality up to now. Due to the dispersion of a tunnel characteristic and due to the Shannon noise description as shown e.g. in [11] causality is not violated.

Around the year 2000 it was observed and calculated that the total tunneling time T equals the reciprocal frequency of the wave entering the barrier $T \approx 1/\nu$ [14, 15]. Data of various experiments are presented in Table 1 showing the overall tunneling time. The various observations and calculations provoked the interest in the three time segments of a tunneling process.

The relation $T \approx 1/\nu$ does not inform about the time $\tau$ spent in the barrier but could be measured by a set-up displayed in Fig. 1, by a high frequency measurement using a network analyzer or by an acoustic experiment at large dimensions -- considering Brillouin's expectation that zero time holds for waves of any field. However already in the first superluminal tunneling experiments it was mentioned that the time inside the barrier may be zero [2].

| Tunneling barriers | T | $T=1/\nu$ |
|---|---|---|
| Undersized waveguide | 130 ps | 115 ps |
| Frustrated total reflection | 117 ps | 120 ps |
| Double prisms | 87 ps | 100 ps |
| Photonic lattice | 2.13 fs | 2.34 fs |
| Photonic lattice | 2.7 fs | 2.7 fs |
| Electron field-emission tunneling | 7 fs | 6 fs |
| Electron ionization tunneling | ≈ 100 as | 105 as[+] |
| Acoustic (phonon) tunneling | 0.8 µs | 1 µs |
| Acoustic (phonon) tunneling | 0.9 ms | 1 ms |
| Neutron (resonant level) | 0.217 µs | 0.236 µs* |
| Neutron (non-resonant) | > 19 ns | 33 ns * |
| Rb Atom (BEC) [++] | 0.62 ms | ≈ 60 ms** |

*Table 1: Results of measured total tunneling time T. Obviously there is a universal tunneling time: $T \approx 1/\nu$; observed over 15 orders of magnitude. Data [++] and * [5]. (Bose-Einstein-Condensation) Data ** are calculated values from the measured tunneling time by the relation $\nu h = kT$. This value is compatible with the barrier height of 135 nK and 180 nK (11).*

The barrier crossing time $\tau = 0$ was first observed by a zero phase shift of the wave between the tunnel in- and output (network analysis) [2], later by the direct time measurement in the double prisms as shown in Fig.1 and further on in a direct pulse – time – delay measurement in the microwave range [11]. In the following table 2 we present measured zero time data of electromagnetic and electron waves. Zero tunneling time was calculated by Low and Mende [1, 18, 19]). However they were obviously frightened by the calculated result having in mind the causality [19]. As mentioned above, the total measured time a signal uses crossing a barrier may be superluminal, the causality is not violated.

**Examples of measured and calculated zero tunneling times $\tau = 0$**

*Microwave phase time (5, 11)*
*Microwave double prisms gap (11)*
*Electron (7*, 1)*
*Ammonia (22)*
*Atom Rb (5)*

*Table 2: data $\tau = 0$*

*The laser field induced tunneling is presumably only caused by a reduction of the potential barrier height and not due to a change of the electron’s basic energy see for instance [16].
The tunneling process was first analyzed by F. Hund [22]. He studied amongst others the inversion oscillation of the ammonia molecule $NH^3$ and obtained a frequency near 24 GHz. For some time this very stable oscillation frequency was applied as a time standard. We conclude that the barrier transition of the inversion of $NH^3$ takes place in zero time in this tunneling process. Ammonia represents another case of Brillouin`s and Schrödinger`s wave mechanics. Zero tunneling time was calculated and discussed for example by F. Low et al. [1, 18, 19].

# Summing up

The study of many different experimental and theoretical zero time barrier tunneling data made us conclude that the tunneling time $\tau$, i.e. the time spent inside a barrier is zero independent of the special potential situation. Our claim is in agreement with many theoretical data on evanescent modes exemplary found in the Ref. [11, 19 - 21] and Brillouin’s assumption that wave mechanics hold for all fields. However, we could not give an answer about the 4th dimension. We have explained the tunneling process to be a virtual one with zero time in the 4th space-time dimension of the Minkowski diagram [21]. In addition, the energy of the tunneling wave is not measurable. The microwave experiments are easy to perform. Two pupils reproducing the double prisms experiment won the first prize in the National German Youth Research Competition 2019 of the state Rheinland Pfalz [23].
Definitively, tunneling represents a non-relativistic process. The Wick Rotation operator may represent a candidate to eliminate real space and time co-ordinates inside the barrier?